\documentclass[aps,pra,epsfigure,twocolumn,longbibliography]{revtex4-1}
\usepackage{dcolumn}    
\usepackage{bm} 
\usepackage{graphicx}
\usepackage{amsmath}    
\usepackage{latexsym}
\usepackage{amsfonts}   
\usepackage{amssymb}
\usepackage{array}      
\usepackage{epsfig}
\usepackage{txfonts}
\usepackage{color}
\usepackage[colorlinks=true,linkcolor=blue,urlcolor=blue,citecolor=blue,pdfusetitle]{hyperref}
\usepackage{hyperref}
\usepackage[normalem]{ulem}

\newcommand{\ket}[1]{\left\vert#1\right\rangle}
\newcommand{\bra}[1]{\left\langle#1\right\vert}

\newcommand{\blah}{blah\\blah\\blah\\blah\\blah.}
\newcommand{\QD}{\text{quantum Darwinism}}

\catcode`\|=\active \def|{
\fontencoding{T1}\selectfont\symbol{124}\fontencoding{\encodingdefault}}

\begin{document}
\title{Collisional unfolding of quantum Darwinism}
       
\author{Steve Campbell,$^{1,2}$ Bar\i\c{s} \c{C}akmak,$^{3,4}$ \"{O}zg\"{u}r E. M\"{u}stecapl{\i}o\u{g}lu,$^3$ Mauro Paternostro,$^5$ and Bassano Vacchini$^{6,1}$}
\affiliation{$^1$Istituto Nazionale di Fisica Nucleare, Sezione di Milano, via Celoria 16, 20133 Milan, Italy\\
$^2$School of Physics, Trinity College Dublin, Dublin 2, Ireland\\
$^3$Department of Physics, Ko\c{c} University, \.{I}stanbul, Sar\i yer 34450, Turkey\\
$^4$College of Engineering and Natural Sciences, Bah\c{c}e\c{s}ehir University, Be\c{s}ikta\c{s}, Istanbul 34353, Turkey\\
$^5$Centre for Theoretical Atomic, Molecular and Optical Physics, School of Mathematics and Physics, Queen's University Belfast, Belfast BT7 1NN, United Kingdom\\
$^6$Dipartimento di Fisica ``Aldo Pontremoli", Universit{\`a} degli Studi di Milano, via Celoria 16, 20133 Milan, Italy }

\begin{abstract}
We examine the emergence of objectivity via quantum Darwinism through the use of a collision model, i.e. where the dynamics is modeled through sequences of unitary interactions between the system and the individual constituents of the environment, termed ``ancillas". By exploiting versatility of this framework, we show that one can transition from a ``Darwinistic" to an ``encoding" environment by simply tuning their interaction. Furthermore we establish that in order for a setting to exhibit quantum Darwinism we require a mutual decoherence to occur between the system and environmental ancillas, thus showing that system decoherence alone is not sufficient. Finally, we demonstrate that the observation of quantum Darwinism is sensitive to a non-uniform system-environment interaction.
\end{abstract}
\date{\today}
\maketitle

\section{Introduction}
According to the framework of open quantum systems, the continuous monitoring of a quantum system by the environment that is coupled to it induces the emergence of classical-like behavior in the latter through the phenomenon of decoherence~\cite{BreuerPetruccione}. Fragile quantum superpositions are spoiled over time, while more robust mixtures of classical-like states~\cite{ZurekRMP} are left to account for the properties of the open system at hand. This notion was made more rigorous by Zurek by introducing the notion of environment induced superselection, or einselection~\cite{ZurekRMP}. Through their mutual interaction, an environment is in effect measuring the system. The essence of einselection is to single out a special ``preferred basis" -- the {\it pointer states} -- that are able to survive this monitoring and therefore remain invariant.  Thus, pointer states are those that remain intact despite their interaction with an environment while, as mentioned above, their superpositions are suppressed~\cite{ZurekRMP}.

In such a picture, the environment (and observers attached to it) typically play a rather passive role, simply embodying sinks for the information leaving the system. Such passivity is epitomized by the technical step embodied by the tracing out of the environmental degrees of freedom, which is routinely performed when looking at the dynamics of the sole system. Yet, much insight into the decoherence process -- and the ensuing transition from quantum to classical description -- can be gathered by promoting the environment (and corresponding observers) to the substantially more active role implied by the consideration of its dynamical nature. By bringing back into the picture the possibility that the interaction between system and environment establishes correlations between them, the framework of quantum Darwinism~\cite{ZurekRMP, KohoutPRA2006, ZurekNatPhys2009} aims at characterizing the process of loss of quantum coherences. 

In the quantum Darwinistic picture, the environment is (arbitrarily) partitioned in disjoint and independent elements $E_j~(j=1,..,N)$, each coupled to the system $S$ via an interaction Hamiltonian $\sum^N_{j=1}H_{SE_j}$. Notice that the elements of the environment do not need to be {\it elementary} per se, and might comprise more than a single particle, for instance. Observers are attached to each element of the environment to acquire information on the state of the system. In this context, the mutual information (MI) $\mathcal{I}_{S\mathcal{E}_{f}}$ shared between the system and a given portion of the environment plays a crucial role in the characterization of the emergence of quantum Darwinism. Specifically, we formally introduce ${\mathcal{I}}_{S\mathcal{E}_{f}}$ as  
\begin{equation}
\label{entropyfunction}
{\mathcal{I}}_{S\mathcal{E}_{f}}=S_S+S_{\mathcal{E}_{f}}-S_{S\mathcal{E}_{f}},
\end{equation}
where $S_{k}=-\text{Tr}[\rho_k\log\rho_k]$ is the von Neumann entropy of the state $\rho_k$ of subsystem $k$ and $\mathcal{E}_f$ denotes a fraction of the environment comprising, in general, of more than a single constituent.

The key observation at the basis of the phenomenology of \QD~is that the system-environment states produced by the decoherence process contain correlations encoded in many copies of classical information about $S$. That is, at the occurrence of the quantum-to-classical transition, the classical information gathered by the environment on the state of $S$ is elevated to the status of an element of objective reality and is thus redundantly encoded in the environment itself~\cite{ZurekNatPhys2009}. Therefore, ${\mathcal{I}}_{S\mathcal{E}_{f}}$ will not depend on the actual size $f$ of ${\cal E}_f$. Such a situation has an important implication on the amount of information accessible by interrogating a portion of the environment. Even with access to only a single environmental unit which has interacted with the system, an observer has, in principle, all the available information about the system and it is not possible to obtain any new information by intercepting more environmental units. While there can be different and more restrictive definitions of objectivity for which \QD~is not sufficient but a necessary condition~\cite{HorodeckiPRA2015, KorbiczPRL2014, CastroPRA2018, Le2018}, \QD~assumes that if the independent measurements of distinct sub-environments by different observers yield the same information on the system then the system is in a classically objective state. The emergence of \QD, and therefore the objectivity from a \QD~perspective, explicitly refers to this redundant information encoding throughout the environment, and is therefore characterized by a plateau emerging at ${\mathcal{I}}_{S\mathcal{E}_{f}}\!=\!S_S$ for any fraction, $f$. In the remainder of this paper we will actually consider the quantity $\bar{\mathcal{I}}_{S\mathcal{E}_{f}}={\mathcal{I}}_{S\mathcal{E}_{f}}/S_S$, i.e. the MI per unit of system-state entropy.

Despite the fact that the topic has been the focus of much attention recently~\cite{ZwolakPRL2009, ZwolakPRA2010, ZurekNJP2012, ZurekSciRep2013, ZwolakPRL2014, ZurekSciRep2016, ZurekPRA2017, PianiNatComms2015, BalaneskovicaEPJD2015, MendlerEPJD2016, GarrawayPRA2017, GiorgiPRA2015, GalveSciRep2016, MauroArXiv, BaldijaoArXiv}, including some seminal experimental investigations~\cite{Brunner2008, Burke2010,PaternostroPRA2018, PanArXiv, ZwolakArXiv}, the phenomenology of the process through which \QD~emerges is still only partially clear or characterized. A full understanding would indeed go together with a complete grasping of decoherence, which is very much a manifestation of the physical mechanism responsible for \QD. In an attempt at rounding our comprehension of its foundations beyond the constraints set by the picture of disjoint and non-interacting environments, recent studies have suggested the detrimental role of non-Markovianity for the manifestation of \QD~\cite{GalveSciRep2016,GiorgiPRA2015,MauroArXiv}. The key merit of such investigations is to have highlighted the relevance of an assessment of the effect that possible intra-environment interactions or correlations might have in the establishment of the Darwinistic phenomenology. Furthermore, recently, the framework laid out by \QD~has also been used in investigating one of the most fundamental discrepancies between quantum and classical theories, namely the violation of non-contextuality inequalities in the former~\cite{BaldijaoArXiv}.

Adding on the list of open questions related to the origin and features of \QD, it is worth mentioning that, recently, its relation with spectrum broadcasting structures has been addressed and studied~\cite{HorodeckiPRA2015, KorbiczPRL2014} to show that certain entangled states may satisfy \QD~without being, strictly speaking, objective at all, the conundrum being the violation of the ``no-perturbation by measurement" condition that is a pre-requisite of objective states. Further elaborations on the discrepancies between the Darwinistic framework and spectral broadcasting structures reveals that the latter imply the former, but not vice-versa, thus calling for a better understanding of the link between \QD, quantum correlations, and broadcasting structures~\cite{CastroPRA2018,Le2018}, which has been investigated by directly addressing physical models for system-environment interactions~\cite{KorbiczPRA2017, LewensteinPRA2017, KorbiczPRL2017, KorbiczPRA2018, CastroPRA2018, KorbiczArXiv}. Thus it is clear that while \QD~provides a promising framework for understanding the emergence of classicality in quantum systems, it is not without its issues. The aim of the present work is to provide a full characterization of \QD~and its role in the emergence of the quantum-to-classical transition by taking a new approach, stemming from the powerful framework embodied by collisional models~\cite{FrancescoQMQM}, to the description of the Darwinistic phenomenology.
 
Collisional models offer a remarkably versatile setting to explore open quantum systems starting from a complete microscopic description \cite{scarani2002, BuzekPRA, BreuerRMP, BreuerRMP, FrancescoQMQM, CiccarelloPRA2013, SantosPRA2012, RuariPRA, StrunzPRA2016, BarisPRA, CampbellPRA2018, EspositoPRX}. In the standard framework the environment is composed of a large, even infinite, number of (generally) finite dimensional constituents termed {\it ancillae}, each with the same initial state. The system then interacts with each ancilla sequentially and, typically, only once, such that after their mutual interaction the ancilla degrees of freedom can be traced over, thus giving rise to the reduced dynamics of the system. We will employ the same basic setting with a few important differences. In order to evaluate Eq.~\eqref{entropyfunction} we will require access to the whole state of the environment, therefore we will fix the number of ancillae to be finite. Additionally, exploiting the versatility of collision models, we allow the system to interact with the same ancilla repeatedly, which provides a possible means to explore non-Markovian dynamics. As commented previously, there have been indications that non-Markovianity hinders \QD~\cite{GiorgiPRA2015, GalveSciRep2016}, however our results indicate that not all manifestations of non-Markovianity prevent classical objectivity from emerging.  We remark that a different form of non-Markovian dynamics which has been extensively studied using collision models relies on allowing for intra-ancilla collision to take place~\cite{RuariPRA, StrunzPRA2016, BarisPRA, CampbellPRA2018}. The consequence of these considerations to \QD~will be the focus of future work.

The remainder of this work is organized as follows. In Sec.~\ref{model} we set the model under consideration, illustrate its features, and introduce the figure of merit that will be used in our analysis. Sec.~\ref{mutual} explores the fundamental mechanisms at the basis of the phenomenology that we reveal, including the role that different forms of quantum correlations have on the emergence of genuine objectivity~\cite{CastroPRA2018}. Sec.~\ref{nonuni} addresses the case of biased system-environment interactions. Finally, in Sec.~\ref{conclusions} we draw our conclusions and leave room for further investigations. 

\section{Quantum Darwinism in a collisional model}
\label{model}

It is well established that the form of the interaction between the system and the environment, as well as the environment's initial state, have a significant effect on whether classical objectivity emerges through \QD~(QD)\cite{ZurekNJP2012}. We will consider a qubit collision model, i.e. the system and all $N$ ancillae are two-level quantum systems. We allow for the general two-body interaction term between system and ancilla
\begin{equation}
\label{interactionterm}
H_{S E_{k}}=\sum_{j=x,y,z}J_j \left(\sigma_S^j\otimes\sigma_{E_k}^j\right)
\end{equation}
such that a single collision corresponds to the application of the unitary evolution operator $U\!=\!e^{-i H_{SE_{k}}t}$ for a fixed time interval, $t$, and where a single ancilla is labeled $E_k$. Furthermore we assume the system and all ancillae begin in a product state
\begin{equation} 
\label{initial}
\ket{\psi_0} =\ket{\phi}_S\bigotimes_{k=1}^N\ket{\Phi}_k,
\end{equation}
with $\ket{\phi}_S=\alpha\ket{\uparrow}+\beta\ket{\downarrow}$ and $\ket{\Phi}_k\!=\!1/\sqrt{2}(\ket{\uparrow}_k+\ket{\downarrow}_k)=\ket{+}_k$ where $\{ \ket{\downarrow}, \ket{\uparrow} \}$ are the eigenstates of their respective free Hamiltonian's, i.e. $H\!=\! \omega\sigma^z$ (throughout we will work in units of $\hbar\omega=1$). The initial state of the environment thus features typical quantum coherences which turn out to be crucial for the onset of QD, however, we remark QD in non-idealized environments has been explored~\cite{ZwolakPRL2009, ZwolakPRA2010}. We consider a discretized evolution such that one step corresponds to a single collision between the system and an ancilla. To comply with the general QD setting, the interactions can generally take place with a randomly chosen ancilla. For a uniform coupling (i.e. an equal number of collisions with each ancilla), this turns out to be equivalent to the standard treatment of collision models in terms of sequential interactions.  

\begin{figure}[t]
{\bf (a)} \hskip4cm {\bf (b)}\\
\includegraphics[width=0.5\columnwidth]{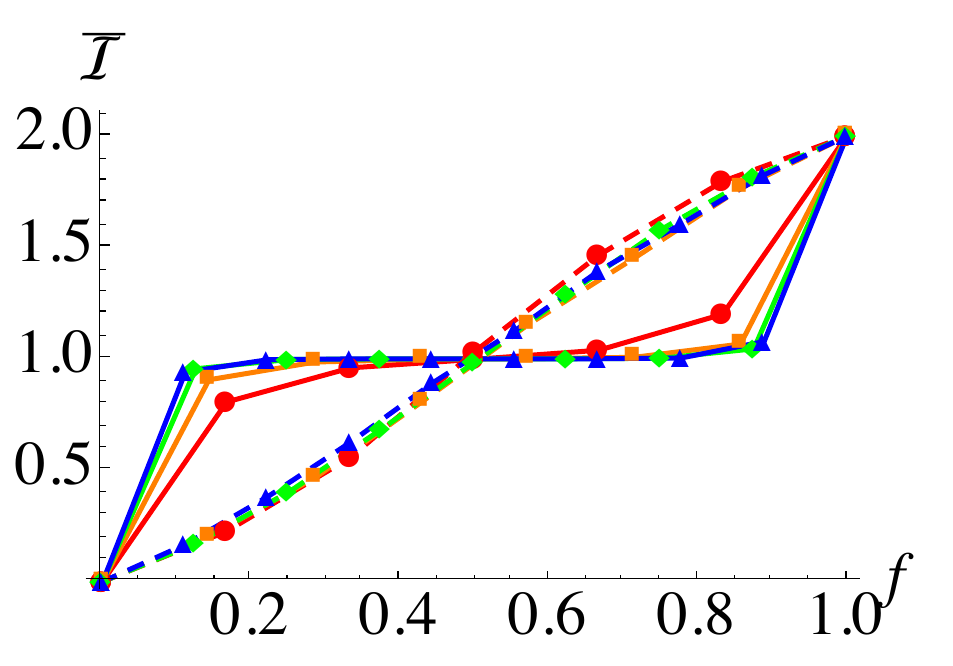}\includegraphics[width=0.5\columnwidth]{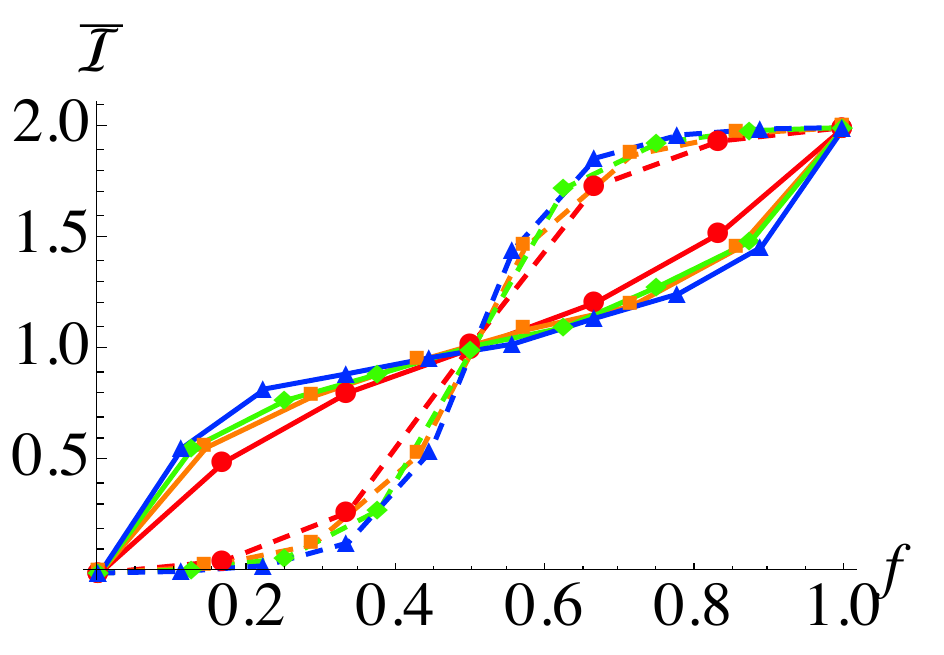}\\
\caption{$\bar{\mathcal{I}}$ as a function of environment fraction, $f$, for $N\!\!=\!\!6$ [red, circles], 7 [orange, squares], 8 [green, diamonds] and 9 [blue, triangles]. We consider two types of system-environment interaction as given in Eq.~\eqref{interactionterm}: the $Z$-interaction [solid curves] where evidence of QD appears, and the $XX$-interaction [dashed curves] where we see an encoding behavior. {\bf (a)} Weak interactions with $J\!=\!1$ and $t\!=\!0.025$. {\bf (b)} Strong interactions with $J\!=\!t\!=\!1$. In both plots the system randomly chooses an environmental ancilla to collide with for a total of $250$ collisions and all ancillae are initially in $\ket{+}$. The initial state of the system is randomly chosen and the plotted curves are the result of averaging over 50 simulations.}
\label{fig1}
\end{figure}
In Fig.~\ref{fig1} we examine two representative types of interaction, the solid curves correspond to $J_x\!=\!J_y\!=\!0$ and $J_z\!=\!J$, i.e. a $Z$-coupling that gives rise to pure dephasing, while the dashed curves correspond to an exchange interaction with $J_x\!=\!J_y\!=\!J$ and $J_z\!=\!0$. In panel {\bf (a)} we consider ``weak" collisions, taking $t\!=\!0.025$, with various sized environments and a total of $250$ collisions take place. Here, as expected, we observe a Darwinistic behavior for the pure dephasing dynamics, i.e. the characteristic plateau emerges at $\bar{\mathcal{I}}\!=\!1$, and the information regarding the state of the system has been redundantly encoded in the ancillae. In panel {\bf (b)} we consider strong interactions with $t\!=\!1$. In this case behavior is significantly less clear cut. While the functional behavior is still qualitatively consistent with QD, evidently there is no sharp plateau appearing. Thus, from the standpoint of objectivity, the information about the system is not completely redundantly encoded, therefore an observer with access to larger portions of the environment has access to more information about the system.

To understand this behavior we recognize that strong interactions mean that the system (or an ancilla) rarely completely decoheres, thus from one collision to the next the amount of coherence in both $S$ and $E_k$ vary significantly. This is a crucial observation that we will return to later.  For the $XX$ interaction, both panels show that there is no redundant encoding of the system information and we do not see any signatures of QD. We find $\bar{\mathcal{I}}$ takes on an $S$-shaped profile indicating an ``encoding" environment, i.e. to learn about the system one requires access to at least half of the environmental ancillae~\cite{ZurekRMP}. In this case, since our ancilla initial state is an eigenstate of $\sigma_x$, the $XX$ model allows for ancilla monitoring of spin components including those in the $x$-direction. As a result, redundant spreading of the system information throughout the environmental units are suppressed. Therefore, it is possible to conclude that choosing the initial state of the total system (system+environment) as an eigenstate of the interaction Hamiltonian between them prevents QD to emerge. These conditions bring the collision model closer to the quantum Zeno regime, the opposite of QD. Qualitatively similar behaviors to that of the $XX$-interaction are found for a general anisotropic Heisenberg interaction Eq.~\eqref{interactionterm}, indicating the special role of the dephasing mechanism in revealing QD. As our focus is on understanding QD, in what follows we we will focus on the dephasing case and leave a more involved the study of the $XX$ and other interaction models for a future work.


\section{Quantum Darwinism as a consequence of mutual dephasing}
\label{mutual}
Finding that classical objectivity emerges when the dynamics of the system is purely dephasing is, in itself, not surprising. However, besides its versatility, the collision model framework allows us to gain significantly more insight into the underlying mechanism as we will now show in a fully analytic manner. For the $Z$-interaction, due to the commutativity between the independent collisions, we can write down the state of the whole system-environment compound as
\begin{equation}
\label{psit}
\ket{\psi}=\alpha\ket{\uparrow}_S\bigotimes_{k=1}^N \ket{\Phi_+}_k +\beta\ket{\downarrow}_S\bigotimes_{k=1}^N \ket{\Phi_-}_k
\end{equation}
with $\ket{\Phi_\pm}_k=e^{\mp i\sum_j g_{j,k}} \left(\ket{\uparrow}_k+e^{\pm2i\sum_j g_{j,k}}\ket{\downarrow}_k\right)/\sqrt2$
where $g_{j,k}$ is the interaction strength (i.e. $J_z t$) of the $j$-th collision between the system with the $k$-th ancilla and therefore the summation characterizes the cumulative strength after all $S$-$E_k$ collisions. It is reasonable to assume that the system collides with all ancillae with equal strength and an equal number of times, $n$, and thus, $\sum_j g_{j,k}\!=\!nJ_zt\!=\!g$ for all $k$. In this regard the model shares the basic features of a spin-star configuration~\cite{SpinStar}. From close examination of Eq.~\eqref{psit}, it becomes apparent that regardless of whether we are interested in the system and/or any given environmental fraction, $\mathcal{E}_f$, the resulting density matrices have at most two non-zero eigenvalues. These are the only quantities necessary to evaluate Eq.~\eqref{entropyfunction}. In what follows we will initialize the system $\ket{\phi}_S=\ket{+}_S$ (we remark this does not affect the generality of our results), and we thus have the concise expressions for the eigenvalues 
\begin{equation}
\lambda^\pm_{k}= \left[1\pm  \cos(2 g)^{\kappa_k}\right]/2~~~~(k=S,{\cal E}_f,S{\cal E}_f),
\end{equation}
with $\kappa_S=N$, $\kappa_{{\cal E}_f}=r$ and $\kappa_{S{\cal E}_f}=\kappa_S-\kappa_{{\cal E}_f}$, where $r$ is the number of ancillae in a given fraction, such that $f=r/N$.

\begin{figure}[t]
{\bf (a)} \hskip0.45\columnwidth {\bf (b)}\\
\includegraphics[width=0.5\columnwidth]{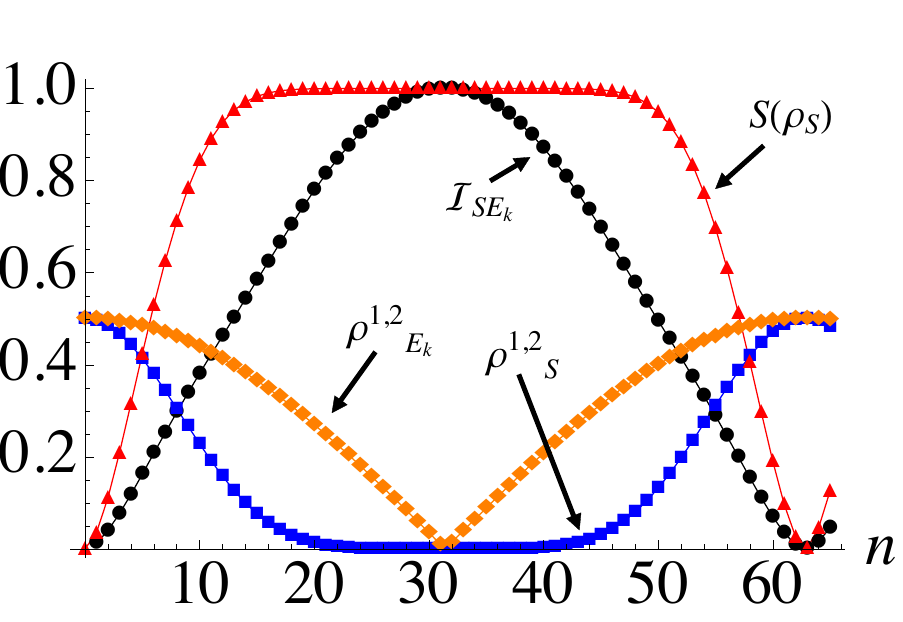}\includegraphics[width=0.5\columnwidth]{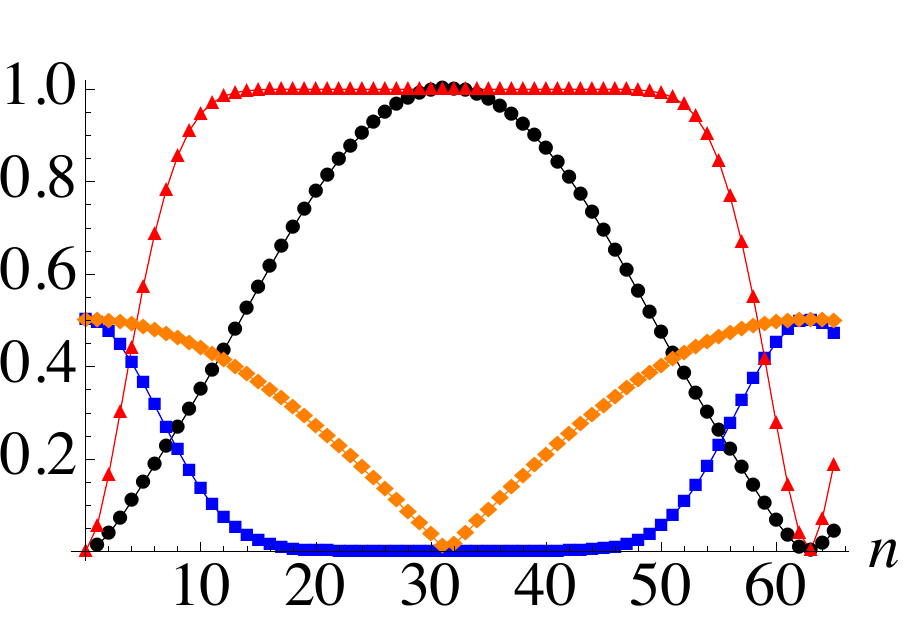}\\
{\bf (c)} \hskip0.45\columnwidth {\bf (d)}\\
\includegraphics[width=0.5\columnwidth]{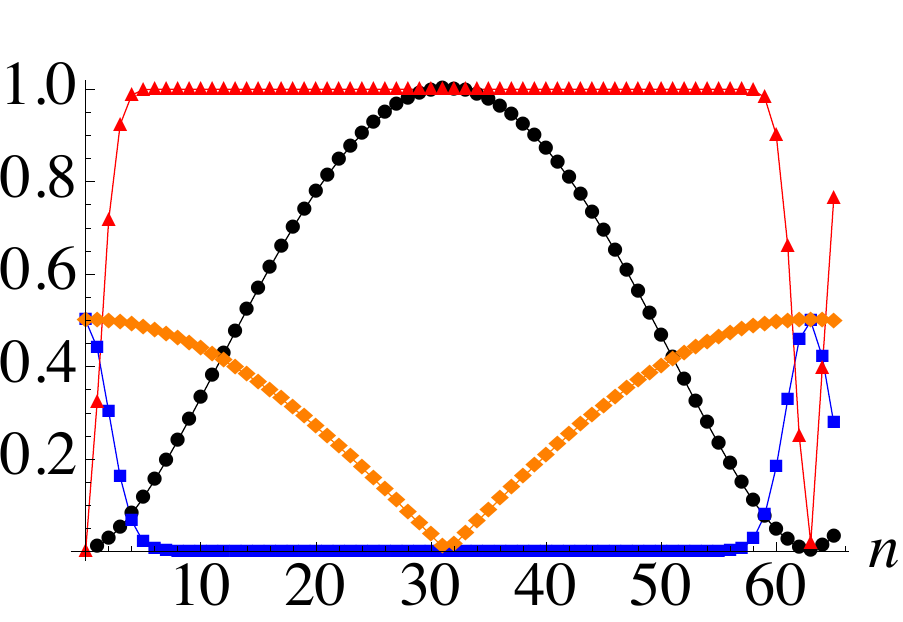}\includegraphics[width=0.5\columnwidth]{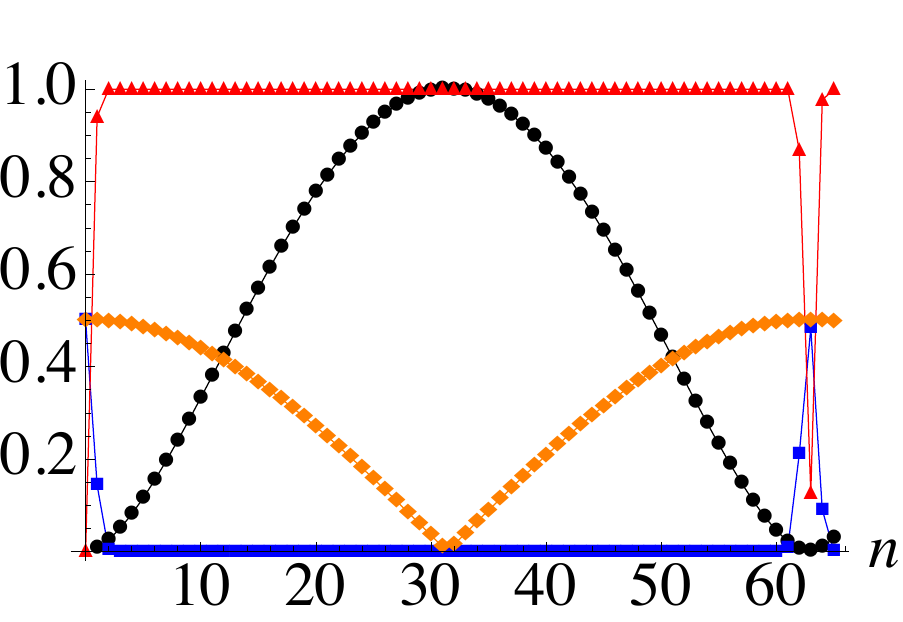}\\
\caption{We consider a system prepared in state $|+\rangle_S$ interacting with a $N$-qubit environment for a dimensionless time $J_zt\!=\!0.025$, so that the dynamics exhibits periodicity for $n\sim\!\!62$. All panels show the MI between the system and a single ancilla $I(\rho_{SE_k})$ [black circles], the entropy of the state of the system $S(\rho_S)$ [red triangles], the coherences in the state of the system $|\rho_S^{1, 2}|$ [blue squares], and those in the state of a single ancilla $|\rho_{E_k}^{1, 2}|$ [orange lozenges]. We consider environments of growing size with {\bf (a)} $N\!=\!6$, {\bf (b)} $N\!=\!10$, {\bf (c)} $N\!=\!100$, and {\bf (d)} $N\!=\!1000$.}
\label{fig2}
\end{figure}

We next explore the emergence of QD in more detail, making reference, in particular, to the basic motivation of QD: how classical objectivity emerges from the system's interaction with the environment that, despite decohering the system, nevertheless leaves its pointer states intact while also ``keeping track" of them. Indeed, in this situation it is clear that we should only expect to see the characteristic plateau in the mutual information provided the system has been fully decohered by its interaction with the environment. While certainly this is true, interestingly by exploiting our collision model we find that this is only a sufficient condition. The decoherence of the system is effectively dictated by how many collisions it undergoes with the various ancillae. Under the assumption that the system collides with all the environmental ancillae an equal number of times, the size of the environment strongly affects the ability of the system to redundantly encode information throughout the environment during its dephasing. We show this in Fig.~\ref{fig2} where we plot $\vert \rho_S^{1,2}\vert$, i.e. the absolute value of the coherence term of the system when $\rho_S(0)\!=\!\ket{+}\!\bra{+}_S$ as a function of the number of collisions $n$ that {\it each} ancilla is involved in. The total number of collisions is thus $n N$. Considering weak interactions again, for small environmental sizes [such as $N=6$ in Fig.~\ref{fig2} {\bf (a)}] we find that each ancilla must collide a relatively large number of times with the system before the latter is fully decohered [we need $\sim\!\!25$ collisions in the case of Fig.~\ref{fig2} {\bf (a)}]. As we increase the size of the environment -- all the way up to $N\!=\!1000$ in Fig.~\ref{fig2} {\bf (d)} -- the number of required collisions is strongly reduced: in Fig.~\ref{fig2} {\bf (d)} two collisions are already sufficient to almost fully decohere the system. If we are to believe that QD will emerge when the system has spread its information throughout the environment, it is reasonable to assume that once the state of the system is fully decohered we would see signatures of QD. To test this it is sufficient to examine if the mutual information shared between the system and any single environmental ancilla is equal to the entropy of the system [cf. Eq.~\eqref{entropyfunction}]. 

The state of the system, a single ancilla, and their joint state are readily obtainable from Eq.~\eqref{psit}. We find that the MI corresponding to the state of the system and a single ancilla is weakly dependent on $N$. In Fig.~\ref{fig2} we show $\mathcal{I}_{SE_{k}}$ [black circles] and $S(\rho_S)$ [topmost curve, red triangles]. Clearly, only when $\sim\!\!31$ collisions have occurred with each ancilla do we find that $\mathcal{I}_{SE_{k}}\!=\!S(\rho_S)$. Note that for the considered dephasing dynamics with the given coupling strength this corresponds to half the periodicity time. This indicates that regardless of the size of the environment, the system must interact with each of the environmental constituents a sufficient number of times before its information is fully and redundantly encoded. Interestingly this corresponds to when the system has fully decohered the ancilla as shown in Fig.~\ref{fig2} where the orange curves correspond to the coherence term of a single ancilla $|\rho_{E_k}^{1,2}|$, which vanishes only when $\mathcal{I}_{SE_{k}}\!=\!S(\rho_S)$. Thus, QD emerges ``perfectly" for any size of the fraction of the environment being considered when the interaction between the latter and the system is sufficient to completely decohere the state of both. 

We examine more closely this effect in Fig.~\ref{fig2bis} {\bf (a)}, where we study the MI between the system and the environment fraction at hand, $\mathcal{E}_f$, for $N\!=\!100$, and $J_zt\!=\!0.025$. If each ancilla undergoes only five collisions with the system, despite this being sufficient to completely decohere $S$, we clearly see that no redundant spreading of the information has occurred, as there is no plateau appearing. This reflects the fact that the environmental fractions all still contain large amounts of coherence, and as a result still share non-classical correlations with the system~\cite{ZurekSciRep2013}. When the number of collisions ensures that the system and ancilla are fully decohered, in this instance $n\!=\!31$, the sharp characteristic plateau appears. In this case, regardless of what sized fraction of the environment an observer can measure, they will have access to the same amount of information about the system. Notice that for other values of $n$ one might conclude that for a {\it sufficiently large} fraction of the environment QD emerges, i.e. in our example for $n\!=\!15$ (55) we see a Darwinistic plateau emerging for fractions $f\!>\!0.1$ (0.35). Notice that the difference between these required fractions can be traced back to the fact that for $n\!=\!15$ all ancillae are more decohered than for $n\!=\!=55$. Furthermore, requirement for an observer to have access to a such a sufficiently large environment fraction is directly related to the complementarity of the classically accessible information and the genuine quantum correlations still shared between $\mathcal{E}_f$ and the system~\cite{ZurekSciRep2013}.

An interesting conclusion that can be drawn from these results is that QD explains the emergence of classical objectivity regardless of the environmental size. Notice that for small environments the requirement for mutual dephasing to almost fully occur for all players is strong, i.e. the system will only fully decohere when the ancillae are also severely decohered. Thus, when environments are small classical objectivity only emerges when the constituents of the environment are effectively also classical. Conversely, as the size of the environment is enlarged, such that we can consider it as a mesoscopic environment compared to the system, then the relative fraction required for classical objectivity is significantly reduced. However, we stress that unless complete mutual dephasing has occurred, perfect classical objectivity, in the sense that even a single environmental ancilla is sufficient to access all the system's information, is not possible.

\section{Nonuniform system-environment coupling and quantum Darwinism}
\label{nonuni}

The collisional framework allows us an additional freedom regarding how strongly a given fraction of the environment interacts with the system. As we have seen previously, for the information on the state of the system to be redundantly encoded, one requires that the system-environment coupling is such that the environmental ancillae are also completely decohered by the interaction. Our collisional model then allows us to ask how QD is affected when the system interacts with a particular fraction of the environment more than the rest. In particular, we consider a situation where the number of collisions between the system and a single ancilla is sufficient such that both are completely decohered, while the interaction of the system with the remaining ancillae only partially decoheres them. To be more concrete, we consider the same setting as before, i.e. both system and ancillae are all initialized in $\ket{+}$ with $N\!=\!6$, we fix $J_z\!=\!1$ and $t\!=\!0.025$ for each collision, and the system will collide with one ancilla $n\!=\!31$ times, while colliding with the remaining five ancillae $n\!=\!60$ times. This results in the fully decohered state of the system and one ancilla, while the remaining ancillae, despite having repeatedly collided with the system, will be almost back in their initial state [cf. the behavior of $|\rho_{E_k}^{1,2}|$ in Fig.~\ref{fig2} {\bf (a)}]. 
\begin{figure}[t]
{\bf (a)} \hskip0.45\columnwidth {\bf (b)}\\
\includegraphics[width=0.5\columnwidth]{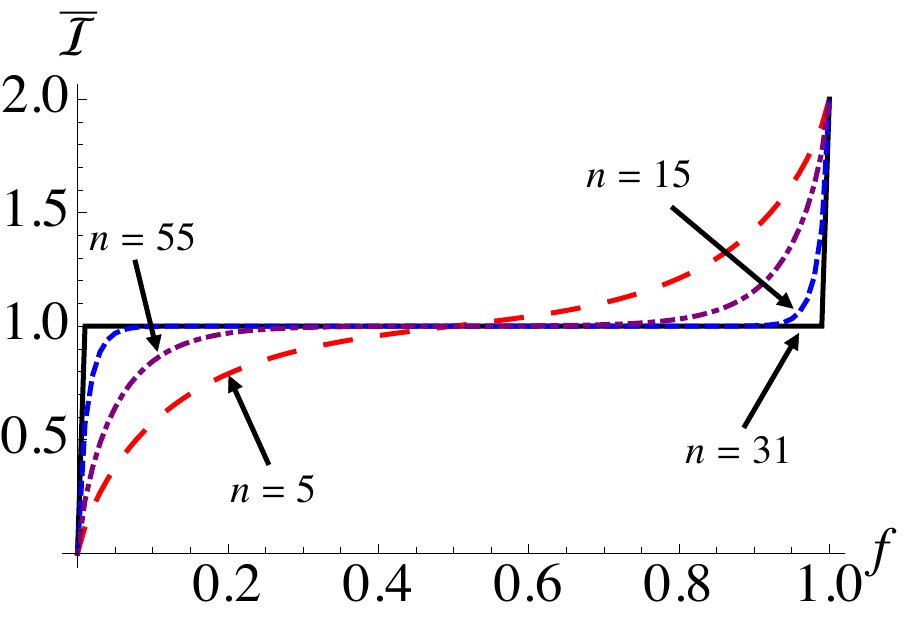}\includegraphics[width=0.5\columnwidth]{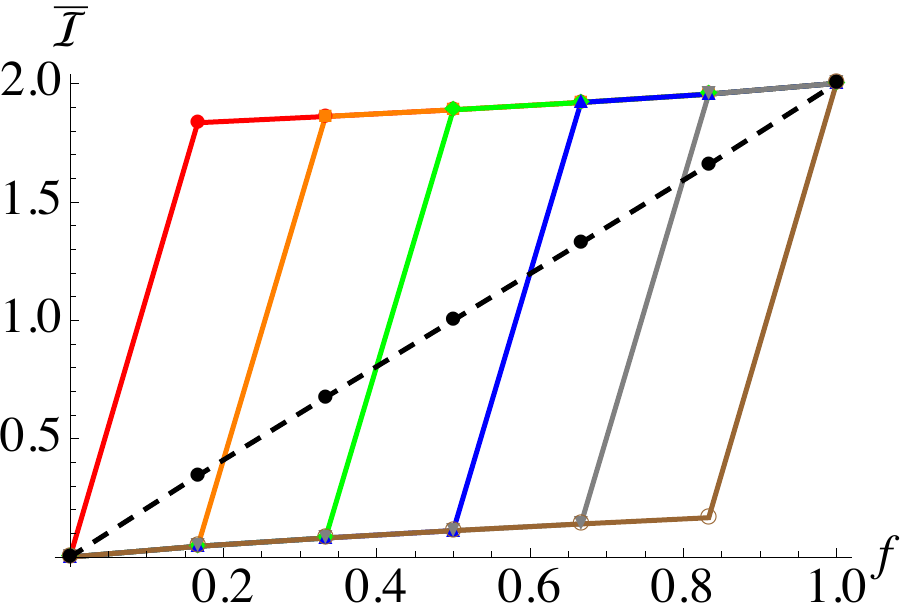}
\caption{{\bf (a)} We show $\bar{\mathcal{I}}$ against the size $f$ of the environmental fraction for $N\!=\!100$. We show curves for various numbers of collisions with each ancilla, such that the total number of system-environment collisions is $100n$. {\bf (b)} We show $\bar{\mathcal{I}}$ against $f$ for an environment consisting of $N=6$ ancillae, each being initially prepared in $\ket{+}_k$. The leftmost red curve corresponds to the case where the system, which is also prepare in $\ket{+}_S$, collides with the first ancilla $n=31$ times, while colliding with the remaining ancillae $60$ times. Each of the other curves from left to right corresponds to a permutation of these collision numbers, with the second curve corresponding to the system colliding with the second ancilla $n=31$ times, while colliding $n=60$ times with all others, and so on. For all such curves we have taken $J_zt\!=\!0.025$. The dashed curve corresponds to the case where the MI is averaged over all possible configurations for a given value of $f$.}
\label{fig2bis}
\end{figure}

After evaluating $\bar{\mathcal{I}}$, we find that the plateau characteristic of Darwinistic behaviors is no longer present. Rather, we find almost all the information about the state of the system is concentrated in a single ancilla, and therefore the amount of information that can be retrieved by examining a portion of the environment is heavily dependent on whether the ancilla that has been fully decohered by the system is contained within the observed fraction. We show this in Fig.~\ref{fig2bis} {\bf (b)}, where the leftmost curve corresponds to the case where the first ancilla in our collisional register undergoes the 31 collisions, while the remaining five undergo $60$ collisions each. An observer who has access to such ancilla can thus learn almost everything about the system. The next curve corresponds to the case where the second ancilla undergoes 31 collisions, while the remaining five undergo $60$ collisions each. This situation entails that an observer with access only to the first ancilla can learn very little about the state of the system. However if they have access to the first two ancillae, the amount of information available to them is almost the maximum possible. Clearly, a similar trend continues when any particular ancilla undergoes 31 collisions. If the observer ignores which ancilla(e) they have access to, such that we average the MI over all possible combinations of $E_k$'s for a given $f$, we still find that QD is lost and the amount of accessible information grows linearly with $f$, as shown by the dashed curve in Fig.~\ref{fig2bis} {\bf (b)}. 

While this is an extreme case of bias in the system-environment interaction, it does raise a subtle point regarding QD and the emergence of objectivity. If we look again at the solid curves in Fig.~\ref{fig1} {\bf (a)}, these correspond to a total of 250 collisions with the ancillae, which is clearly not a multiple of any of the considered environmental sizes. Therefore, in all curves at least one ancilla will have undergone more collisions than the rest. While the behavior here is clearly Darwinistic, there are slight deviations for fraction sizes $f\!=\!1/N$ and $f\!=\!(N-1)/N$, similar to those observed and discussed in Refs.~\cite{ZurekSciRep2013, ZwolakArXiv}. Thus we should caveat the results of Fig.~\ref{fig2bis} {\bf (b)}: if the system interacts almost uniformly with all constituents in the environment, then one will see QD emerging provided this interaction is sufficient to decohere both the system and all ancillae. However, we see the breakdown in QD if the system fully decoheres a small subset of the environmental degrees of freedom, while leaving significant coherences in the remaining fraction despite having repeatedly interacted with them. \bigskip

\section{Conclusions}
\label{conclusions}
In this work we have examined the mechanism that leads to objectivity through \QD~(QD) using a collision model framework. The collision model setting allowed us to show that the nature of the interaction between system and ancillae dictates if QD is observed and one can easily transition from a ``redundantly encoded" to an ``encoding" environment simply by varying this interaction term. Furthermore, we have shown that the decoherence of the system by the environment is only a sufficient condition for QD and for objectivity to emerge we also require the environmental ancillae to be decohered. Finally, exploiting the versatility of collision models, we have shown that a non-uniform system-environment coupling can lead to a complete loss of any signatures of QD. Our work therefore highlights some subtle features that underpin the QD paradigm. Indeed in light of our results one might be inclined to conclude that QD is a rare phenomenon. However, our work rather highlights that under arguably well-justified constraints, namely a dephasing interaction and relatively uniform coupling, we find QD generically emerge. We expect our results to provide a promising avenue with which QD, and other approaches to the emergence of classical objectivity, can be more rigorously tested using collision models where the effects of other phenomena, e.g. non-Markovianity, can be readily introduced and studied, ultimately contributing to understanding the quantum-to-classical transition.

\acknowledgements
SC gratefully acknowledges the Science Foundation Ireland Starting Investigator Research Grant ``SpeedDemon" (No. 18/SIRG/5508) for financial support. MP, B\c{C} and \"{O}EM acknowledge support from Royal Society Newton Mobility Grant (Grant No. NI160057). B\c{C} thanks Queen's University Belfast and the University of Milan for hospitality during the development of part of this work. BV acknowledges support from the EU Collaborative Project QuProCS (Grant Agreement No. 641277) and FFABR. \"{O}EM acknowledges support by TUBITAK (Grant No. 116F303). MP acknowledges support from the DfE-SFI Investigator Programme (Grant No. 15/IA/2864) and the H2020 Collaborative Project TEQ (Grant Agreement No. 766900). \"{O}EM and MP are partially supported by the COST Action "QTSpace" (CA15220).

\bibliography{quantum_darwinism}

\end{document}